\documentclass[10pt,conference,review]{IEEEtran}
\IEEEoverridecommandlockouts

\usepackage[font=small,labelfont=bf,tableposition=top]{caption}
\usepackage{hyperref}
\usepackage{boldline}
\usepackage{color,colortbl}
\usepackage{bigstrut}
\usepackage[ruled]{algorithm2e}
\usepackage{amsmath,amsfonts}
\usepackage{amsmath}
\usepackage{array}
\usepackage{algorithmic}
\usepackage{balance}
\usepackage{blindtext}
\usepackage{booktabs}
\usepackage{caption}
\usepackage{hyperref}
\usepackage[noabbrev]{cleveref}
\usepackage{cite}
\usepackage{comment}
\usepackage{enumitem}
\usepackage{eqparbox}
\usepackage{fancybox}
\usepackage{fancyvrb}
\usepackage{framed}
\usepackage{graphicx}
\usepackage{ifthen}
\usepackage{listings}
\usepackage{mathrsfs}
\usepackage{mdwmath}
\usepackage{mdwtab}
\usepackage{multirow}
\usepackage{pifont}
\usepackage{stfloats}
\usepackage{textcomp}
\usepackage{times}
\usepackage{url}
\usepackage{xspace}
\usepackage[normalem]{ulem}
\usepackage[framemethod=TikZ]{mdframed}
\usepackage{caption}
\usepackage{subcaption}
\usepackage{tabularx}
\usepackage[switch]{lineno}

\usepackage{titlesec}

\usepackage{tcolorbox}
\tcbuselibrary{listings,skins}
 
\usepackage{mathtools}
\usepackage{blindtext}
\usepackage{lstautogobble}

\DeclareGraphicsExtensions{.pdf,.jpeg,.png}
\graphicspath{{figures/}}

\titlespacing\section{0pt}{8pt plus 4pt minus 2pt}{4pt plus 2pt minus 2pt}

\newcommand{\rom}[1]{\uppercase\expandafter{\romannumeral #1\relax}}

\newcommand{\etal}{\hbox{\emph{et al.}}\xspace}
\newcommand{\eg}{\hbox{\emph{e.g.,}}\xspace}
\newcommand{\ie}{\hbox{\emph{i.e.,}}\xspace}

\newcommand{\wrt}{\hbox{\emph{w.r.t.}}\xspace}

\newcommand{\etc}{\hbox{\emph{etc.}}\xspace}

\definecolor{gray50}{gray}{.5}
\definecolor{gray40}{gray}{.6}
\definecolor{gray30}{gray}{.7}
\definecolor{gray20}{gray}{.8}
\definecolor{gray10}{gray}{.9}
\definecolor{gray05}{gray}{.95}
\definecolor{gray01}{gray}{.97}

\newlength\Linewidth
\def\findlength{\setlength\Linewidth\linewidth
\addtolength\Linewidth{-4\fboxrule}
\addtolength\Linewidth{-3\fboxsep}
}
\newenvironment{examplebox}{\par\begingroup
  \setlength{\fboxsep}{3pt}\findlength
  \setbox0=\vbox\bgroup\noindent
  \hsize=0.90\linewidth
  \begin{minipage}{0.90\linewidth}\normalsize}
    {\end{minipage}\egroup
    \textcolor{gray20}{\fboxsep1.2pt\fbox
     {\fboxsep5pt\colorbox{gray05}{\normalcolor\box0}}}
    \endgroup\par\noindent
    \normalcolor\ignorespacesafterend}

\newcounter{RQACounter}

\usetikzlibrary{shadows}
\usepackage{graphics}
\newmdenv[
    tikzsetting= {fill=blueish},
    skipabove=0.33em,
    skipbelow=0.33em,
    linewidth=1pt,
    innerleftmargin=4pt,
    innerrightmargin=4pt,
    innertopmargin=2pt,
    innerbottommargin=2pt,
    linecolor=gray95,
    roundcorner=2pt, 
    shadow=true,
    shadowsize=4pt,
    shadowcolor=gray95
]{questionbox}

\newmdenv[
    tikzsetting= {fill=greenish},
    skipabove=0.33em,
    skipbelow=0.33em,
    linewidth=1pt,
    innerleftmargin=4pt,
    innerrightmargin=4pt,
    innertopmargin=2pt,
    innerbottommargin=2pt,
    linecolor=gray95,
    roundcorner=2pt, 
    shadow=true,
    shadowsize=4pt,
    shadowcolor=gray95
]{answerbox}

\newmdenv[
    skipabove=0.33em,
    skipbelow=0.33em,
    innerleftmargin=4pt,
    innerrightmargin=4pt,
    innertopmargin=2pt,
    innerbottommargin=2pt,
]{lessonbox}

\usepackage{tikz}

\newenvironment{result}
{\begin{answerbox}}
{\end{answerbox}}

\newenvironment{question}
{\begin{questionbox}}
{\end{questionbox}}

\definecolor{javared}{rgb}{0.6,0,0} 
\definecolor{javagreen}{rgb}{0.25,0.5,0.35} 
\definecolor{javapurple}{rgb}{0.5,0,0.35} 
\definecolor{javadocblue}{rgb}{0.25,0.35,0.75} 

\lstdefinestyle{basejava}{
  language=java,
  showstringspaces=false,
  basicstyle=\small\ttfamily,
  keywordstyle=\bfseries\color{javapurple},
  commentstyle=\itshape\blue,
  identifierstyle=\blue,
  frame=none,
  backgroundcolor=\color{white},
}

\lstdefinestyle{CustomJava}{
  belowcaptionskip=\baselineskip,
  breaklines=true,
  xleftmargin=\parindent,
  language=java,
  showstringspaces=false,
  basicstyle=\scriptsize\ttfamily,
  keywordstyle=\bfseries\color{javapurple},
  commentstyle=\itshape\blue,
  identifierstyle=\blue,
  belowskip=1pt,
  frame=shadowbox,
  backgroundcolor=\color{gray01},
  gobble=0
}

\lstdefinestyle{codit}{
  belowcaptionskip=\baselineskip,
  breaklines=true,
  language=java,
  showstringspaces=false,
  basicstyle=\scriptsize\ttfamily,
  keywordstyle=\bfseries\color{javapurple},
  commentstyle=\itshape\blue,
  identifierstyle=\blue,
}

\newtcblisting{mylisting}[2][]{
    arc=3mm,
    listing only, 
    listing style=codit,
    title=#2,
    #1
    }

\newcommand\red[1]{\textcolor[rgb]{1.00,0.00,0.00}{#1}}

\newcommand\blue[1]{\textcolor[rgb]{0.00,0.00,1.00}{{#1}}}

\newcommand\dkgreen[1]{\textcolor[rgb]{0.0,0.6,0}{\textbf{#1}}}

\definecolor{blueish}{RGB}{250, 250, 255}
\definecolor{greenish}{RGB}{250, 255, 250}

\definecolor{gray05}{gray}{0.95}
\definecolor{gray08}{gray}{0.92}
\definecolor{gray10}{gray}{0.90}
\definecolor{gray12}{gray}{0.88}
\definecolor{gray15}{gray}{0.85}
\definecolor{gray18}{gray}{0.82}
\definecolor{gray20}{gray}{0.80}
\definecolor{gray25}{gray}{0.75}
\definecolor{gray30}{gray}{0.70}
\definecolor{gray35}{gray}{0.65}
\definecolor{gray40}{gray}{0.60}
\definecolor{gray45}{gray}{0.55}
\definecolor{gray50}{gray}{0.50}
\definecolor{gray55}{gray}{0.45}
\definecolor{gray60}{gray}{0.40}
\definecolor{gray65}{gray}{0.35}
\definecolor{gray70}{gray}{0.30}
\definecolor{gray75}{gray}{0.25}
\definecolor{gray80}{gray}{0.20}
\definecolor{gray85}{gray}{0.15}
\definecolor{gray90}{gray}{0.10}
\definecolor{gray95}{gray}{0.05}

\newcommand{\tool}{\textsc{Modit}\xspace}

\xspace%

\newcommand{\sts}{$\mathcal{S}2\mathcal{S}$\xspace}

\newcommand{\Comment}[1]{}

\newcommand{\rayb}[1]{{\todo{Baishakhi:  {\color{violet} #1}}}}

\newcommand{\linecode}[1]{\lstinline[escapechar=@,basicstyle=\ttfamily]{#1}~}
\newcommand{\linecodewospace}[1]{\lstinline[escapechar=@,basicstyle=\ttfamily]{#1}}

\newcommand{\cross}{\ding{55}}
\newcommand{\tick}{\ding{51}}

\newcommand{\codit}{\textsc{Codit}\xspace}
\newcommand{\citep}[1]{\cite{#1}}

\newcommand{\pcode}{$C_p$\xspace}
\newcommand{\ncode}{$C_n$\xspace}
\newcommand{\guidance}{$\mathcal{G}$\xspace}
\newcommand{\buggy}{$e_p$\xspace}
\newcommand{\fixed}{$e_n$\xspace}
\newcommand{\context}{$C$\xspace}

\newtcbox{\inlinebox}[1][]{enhanced,
 box align=base,
 nobeforeafter,
 colback=blueish,
 size=small,
 left=0pt,
 right=0pt,
 boxsep=2pt,
 #1}

\newcommand{\highlight}[1]{%
{\footnotesize%
\inlinebox{#1}%
}}

\newcommand{\sdata}{
{$B2F_{s}$}\xspace
}
\newcommand{\mdata}{
{$B2F_{m}$}\xspace
}

\newcommand{\U}{\textbf{\_}}

\newcommand{\rqa}{How accurately does \tool generate edited code \wrt other techniques?}
\newcommand{\rqb}{What  are  the  contribution  of  different  input  modalities in  \tool’s  performance?}
\newcommand{\rqc}{What  is  the  best  strategy  to  encode  multiple input  modalities?}

\newcommand{\RQrepeat}[2]{%
    \begin{question}{
    \small
    \noindent\textbf{RQ#1.~#2}
    }\end{question}
}

\newcommand{\RQ}[2]{%
    {
    \newline
    \refstepcounter{RQACounter} \label{rq-#1}
    \noindent\textbf{RQ\arabic{RQACounter}.~#2}
    \smallskip
    }
}

\newcommand{\RS}[2]{%
    \begin{result}
        \textbf{Result {\ref{rq-#1}}:~}{\emph {#2}}%
    \end{result}
}

\renewcommand{\cref}[1]{\Cref{#1}}

\newcommand{\transBase}{{\em Transformer-base}\xspace}
\newcommand{\transLarge}{{\em Transformer-large}\xspace}

\usepackage[turnoff]{notes}
\IEEEtriggeratref{54}
\begin{document}

\title{On Multi-Modal Learning of Editing Source Code}

\author{
    \IEEEauthorblockN{Saikat Chakraborty}
    \IEEEauthorblockA{
        {Department of Computer Science} \\
        {Columbia University}\\
        New York, NY, USA \\
        \href{saikatc@cs.columbia.edu}{saikatc@cs.columbia.edu}
    }
    \and
    \IEEEauthorblockN{Baishakhi Ray}
    \IEEEauthorblockA{
        {Department of Computer Science} \\
        {Columbia University}\\
        New York, NY, USA \\
        \href{rayb@cs.columbia.edu}{rayb@cs.columbia.edu}
    }
}
\maketitle
\begin{abstract}
In recent years, Neural Machine Translator (NMT) has shown promise in automatically editing source code. Typical NMT based code editor only considers the code that needs to be changed as input and suggests developers with a ranked list of patched code to choose from - where the correct one may not always be at the top of the list. While NMT based code editing systems generate a broad spectrum of plausible patches, the correct one depends on the developers' requirement and often on the context where the patch is applied. Thus, if developers provide some hints, using natural language, or providing patch context, NMT models can benefit from them. 

As a proof of concept, in this research, we leverage three modalities of information: edit location, edit code context, commit messages (as a proxy of developers' hint in natural language) to automatically generate edits with NMT models. To that end, we build \tool, a multi-modal NMT based code editing engine. With in-depth investigation and analysis, we show that developers' hint as an input modality can narrow the search space for patches and outperform state-of-the-art models to generate correctly patched code in top-1 position. 
\end{abstract}

\begin{IEEEkeywords}
Source Code Edit, Neural Networks, Automated Programming, Neural Machine Translator, Pretraining, Transformers
\end{IEEEkeywords}

\section{Introduction}
\label{sec:intro}


Programmers often develop software incrementally, adding gradual changes to the source code. In a continuous software development environment, programmers modify their source code for various reasons, including adding additional functionality, fixing bugs, refactoring, etc. 
It turns out that many of these changes follow repetitive edit patterns~\cite{ray2014uniqueness, meng2011systematic, meng2013lase} resulting in a surge of research effort to automatically generate code-changes learned from past  examples~\cite{meng2011systematic, meng2013lase, rolim2017learning, tufano2019learning, chakraborty2020codit}. 

In particular, Neural Machine Translation (NMT) models have been successful in learning automatic code changes~\cite{chakraborty2020codit, tufano2019learning, tufano2018nmt_bug_fix, tufano2019empirical, jiang2021cure, lutellier2020coconut, chen2019sequencer}. At the core, these models contain an encoder and a decoder --- the encoder encodes the code that needs to be edited, and the decoder sequentially generates the edited code. Such NMT models are trained with a large corpus of previous edits to learn generic code change patterns. In the inference time, given a code fragment that needs to be edited, a trained NMT model should automatically generate the corresponding edited code. 

\begin{figure}[!tbh]
\small
\begin{subfigure}{0.92\linewidth}
\begin{tabular}{l}
\lstset{escapechar=~,style=CustomJava}
\begin{lstlisting}
~\textbf{//Guidance: use LinkedList and fix sublist problem ... }~ 
public void addPicture (String picture){ 
    if ((pictures) == null) {
~\red{\textbf{-}}~       ~\red{\textbf{pictures = new ArrayList<>();}}~
~\dkgreen{\textbf{+}}~       ~\dkgreen{\textbf{pictures = new LinkedList<>();}}~ //correct patch
~\dkgreen{\textbf{+}}~       ~\blue{\textbf{\underline{pictures = new HashSet<>();}}}~ //plausible  patch
    }
    pictures.add(picture); 
}
\end{lstlisting}
\end{tabular}
\end{subfigure}
\caption{\footnotesize \bf Example of an identical code (marked in~\red{red}) changed in two different ways (~\dkgreen{green} and~\blue{\underline{blue}}) in two different contexts, where both can be correct patches. However, based on developers' guidance (top line) to fix a list related problem, ~\dkgreen{green} is the correct patch in this context.}
\label{fig:multiple_ways_edit_example}
\end{figure}








However, learning such generic code changes is challenging. 
A programmer may change an identical piece of code in different ways in two different contexts, both can  potentially be correct patches (see~\Cref{fig:multiple_ways_edit_example}). 
For example, an identical code fragment {\tt pictures = new ArrayList$<>$()} was changed in two different ways: {\tt pictures = new HashSet$<>$();} and {\tt pictures = new LinkedList$<>$()} in two different code contexts. 
Without knowing the developers' intention and the edit context, the automated code editing tools have no way to predict the most intended patches. For instance, in the above example, {\tt LinkedList} was used to fix a sublist-related problem. Once such an intention is known, it is easy to choose a LinkedList-related patch from the alternate options. Thus, such an additional modality of information can 
reinforce the performance of 
automated code-editing tools.

\begin{figure}[!tbh]
\centering

\includegraphics[width=\linewidth]{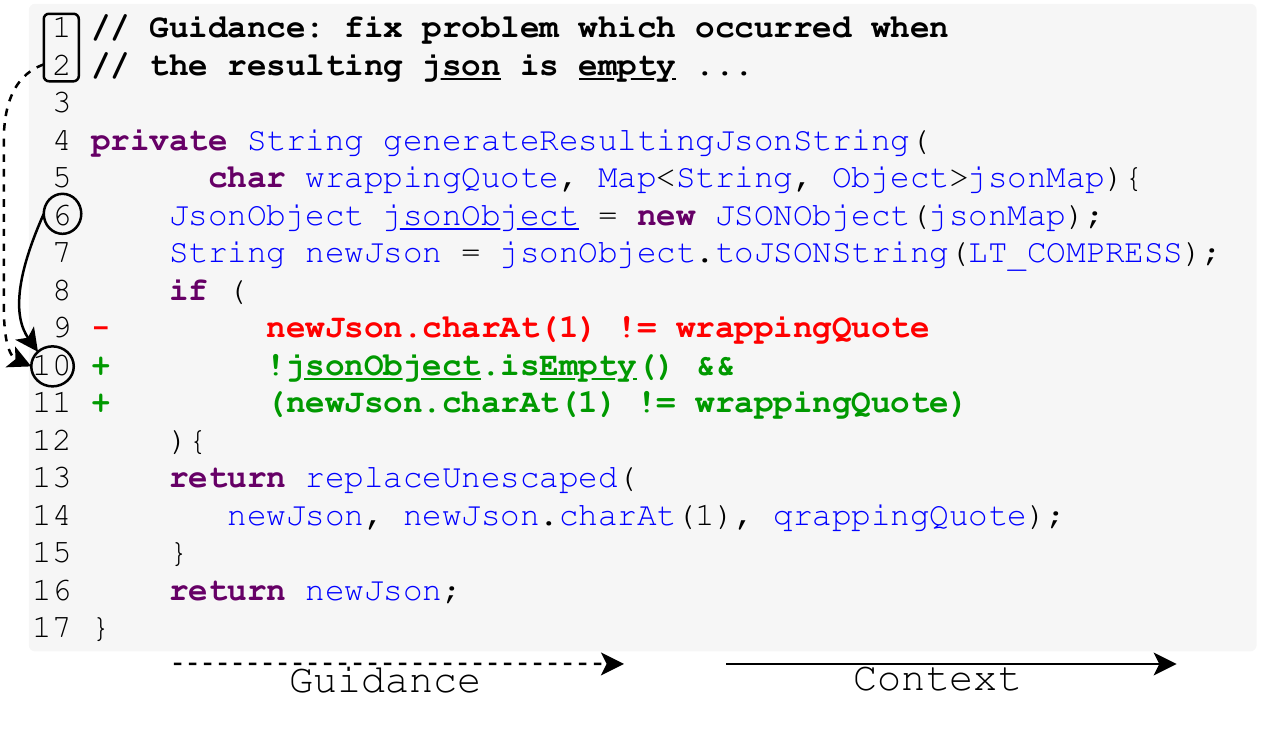}
\caption{\textbf{\footnotesize A motivating example. The guidance provides a brief summary of what needs to be changes. The underlined tokens are directly copied from guidance and context into the patched code.}}
\label{fig:motivating}
\end{figure}

In fact, given just a piece of code without any additional information, it is perhaps unlikely that even a human developer can comprehend how to change it. Consider another real-life example shown in \Cref{fig:motivating}. If a programmer  {\em only} considers the edited expression in line 9, 
it is difficult to decide how to modify it. However, with additional information modalities -- \ie the guidance (line 1,2) and the context (the whole method before the patch), the correct patch often becomes evident to the programmer since the guidance effectively summarises how to change the code and the context provides necessary ingredients for generating a concretely patched code. We hypothesize that such multi-modal information could be beneficial to an automated code-editing tool. To that end, we design \tool, a multi-modal code editing engine that is based on three information modalities: (i) the code fragment that needs to be edited, (ii) developers' intention written in natural language, and (iii) {
explicitly given edit context}. 

In particular, 	\tool is based on a transformer-based~\cite{vaswani2017attention} NMT model. 
As input, \tool takes the code that needs to be edited (\eg the lines that need to be patched), 
additional guidance describing developers' intent, and the context of the edits that are explicitly identified by the developer (\eg the surrounding method body, or surrounding lines of code, etc.). {Note that previous works~
\cite{chakraborty2020codit, chen2019sequencer}
also provided context and the edit location while generating edits; however, they are fed together to the model as a unified code element. Thus, the model had the burden of identifying the edit location and then generating the patch. In contrast, isolating the context from the edit location and feeding them to the model as different modalities provides \tool with additional information about the edits.}

Curating developers' intent for a large number of edits that can train the model is non-trivial. As a proof of concept, we leverage the commit messages associated with the edits to simulate developers' intent automatically. We acknowledge that commit messages could be noisy and may not always reflect the change summary~\cite{Liu2018NMTCommit}. 
Nonetheless, our extensive empirical result shows that, even with such noisy guidance, \tool performs better in generating correctly edited code.

Being a model that encodes and generates source code, \tool needs to both clearly understand and correctly generate programming languages (PL). While several previous approaches~\cite{chakraborty2020codit, wang2020modular} designed sophisticated tree/grammar-based models to embed the knowledge of PL into the model, the most recent transformer-based approaches~\cite{feng2020codebert, guo2020graphcodebert, ahmad2021unified} showed considerable promise with pre-training with a large volume of source code. Since these models are pre-trained with billions of source code written by actual developers, and transformers are known to learn distant dependencies between the nodes, these models can learn about code structures during the pre-training step. Among such pre-trained models,  PLBART~\cite{ahmad2021unified} learns jointly to understand and generate source code and showed much promise in generative tasks. Thus, we chose PLBART as the starting point to train \tool, i.e., we initialize \tool's model with learned parameters from PLBART.  

We evaluate \tool on two different datasets (\sdata, and \mdata) proposed by Tufano~\etal~\cite{tufano2019empirical} consisting of an extensive collection of bug-fix commits from GitHub. 
Our empirical investigation shows that a summary of the change written in natural language as additional guidance from the developer improves \tool's performance by narrowing down the search space for change patterns. The code-edit context, presented as a separate information modality, helps \tool to generate edited code correctly by providing necessary code ingredients (\eg variable names, method names, \etc). \tool generates $\sim$3.5 times more correct patches than \codit showing that \tool is robust enough to learn PL syntax implicitly. Furthermore, \tool generates two times as many correct patches as a large transformer model could generate. 

Additionally, our empirical investigation reveals that when we use one encoder to encode all information modalities rather than learning from individual modalities separately, the model learns representation based on inter-modality reasoning. In contrast, a dedicated encoder for each individual modality only learns intra-modality reasoning. Our experiment shows that a multi-modal/single-encoder model outperforms multi-modal/multi-encoder model by up to 46.5\%.

We summarize our main contributions in this paper as follows. 
\begin{itemize}
    \item We propose \tool -- a novel multi-modal NMT-based tool for automatic code editing. Our extensive empirical evaluation shows that Automatic Code Editing can be vastly improved with additional information modalities like code context and developer guidance. 
    \item We empirically investigate different design choices for \tool. We provide a summary of the lessons that we learned in our experiments. We believe such lessons are valuable for guiding future research. 
    \item We prototype and build \tool and open-source all our code, data in \href{https://git.io/JOudU}{https://git.io/JOudU}.
\end{itemize}

\section{Background}
\label{sec:background}

\subsection{Neural Machine Translation}
\label{background/nmt}

Neural Machine Translation(NMT)~\cite{bahdanau2014neural} is a very well studied field, which has been very successful in translating a sentence from one language to another. At a very high level, input to an NMT model is a sentence ($X = x_1, x_2, ..., x_n$), which is usually a sequence of tokens ($x_i$), and the output is also a sentence ($Y = y_1, y_2, ..., y_m)$ -- sequence of tokens ($y_i$). While learning to translate from $X$ to $Y$, NMT models learn conditional probability distribution $P(Y|X)$
. Such probability distributions are learned \wrt model parameters $\theta$, where model training process optimizes $\theta$ in such a way that maximizes the expected probability distribution of a dataset. An NMT model usually contains an encoder and a decoder. The encoder processes, understands, and generates vector representations of the input sentence. The decoder starts after the encoder and sequentially generates the target sentence by reasoning about the encoder-generated input representation. While sequentially generating the target sentence, the decoder usually performs different heuristic searches (for instance, beam search) to balance exploration and exploitation. 

In recent few years, Software Engineering has seen a wide spectrum of adaptation of NMT. Some prominent application of NMT is SE include Program Synthesis~\cite{yin2017syntactic}, Code summarization~\cite{wei2019code, ahmad2020summarization}, Edit summarization~\cite{Liu2018NMTCommit}, Code Edit Generation~\cite{tufano2019empirical, tufano2019learning, chakraborty2020codit}, Automatic Program Repair~\cite{lutellier2020coconut, jiang2021cure, chen2019sequencer}, etc. These research efforts capitalize on NMTs' capability to understand and generate complex patterns and establish NMT as a viable tool for SE-related tasks.

\subsection{Transformer Model for Sequence Processing}
\label{background/transformer}

Transformer~\cite{vaswani2017attention} model revolutionized sequence processing with attention mechanism. Unlike the traditional RNN-based model where input tokens are processed sequentially, the transformer assumes soft-dependency between each pair of tokens in a sequence. Such dependency weights are learned in the form of attention weights based on the task of the transformer. While learning the representation of a token, the transformer learns to attend to all the input tokens. From a conceptual point of view, the transformer converts a sequence to a complete graph\footnote{\href{https://en.wikipedia.org/wiki/Complete_graph}{https://en.wikipedia.org/wiki/Complete\_graph}}, where each node is a token. The weights of the edges are attention weights between tokens which are learned based on the task of the transformer. The transformer encodes each token's position in the sequence (positional encoding) as part of the input. In such a way, the transformer learns long-range dependency. Since its inception, the transformer is very successful in different NLP understanding and generation tasks. Transformers' ability of reasoning about long-range dependency is proved  useful for several source code processing task including code completions~\cite{kim2020code}, code generation~\cite{svyatkovskiy2020intellicode}, code summarization~\cite{ahmad2020summarization}.

\subsection{Transfer Learning for Source Code}
\label{background/transfer-learning}


In recent few years, Transfer learning shows promise for a wide variety of SE tasks. Such transfer learning aims at learning task agnostic representation of source code and reuse such knowledge for different tasks. One way to learn such task agnostic representation of input is pre-training a model with a large collection of source code. The learning objective of such pre-training is often understanding the code or generating the correct code. A pre-trained model is expected to embed the knowledge about source code through its parameters. Such pre-trained models are later fine-tuned for task-specific objectives. CuBERT~\cite{kanade2019pre}, CodeBERT~\cite{feng2020codebert}, GraphCodeBERT~\cite{guo2020graphcodebert} are all transformer-based encoder models which are pre-trained to understand code. Such models are primarily trained using Masked Language Model~\cite{dong2019unified}, replaced token prediction~\cite{feng2020codebert}, semantic link prediction~\cite{guo2020graphcodebert}, etc. 

For code generation, CodeGPT~\cite{CodeXGLUE, jiang2021cure} pre-trains a transformer-based model to generate general-purpose code sequentially. More recently, PLBART~\cite{ahmad2021unified} pre-trained transformer-based model jointly for understanding and generating code with denoising auto-encoding~\cite{lewis2019bart}. PLBART consists of an encoder and a decoder. The encoder is presented with slight noise (for instance, token replacement) induced code, and the decoder is expected to generate noise-free code. Since code editing task requires both the understanding of code and code generation, we chose PLBART as the base model for \tool.

\section{\tool}
\label{sec:tool}

\begin{figure*}[!htb]
    \centering
    \includegraphics[width=0.95\linewidth]{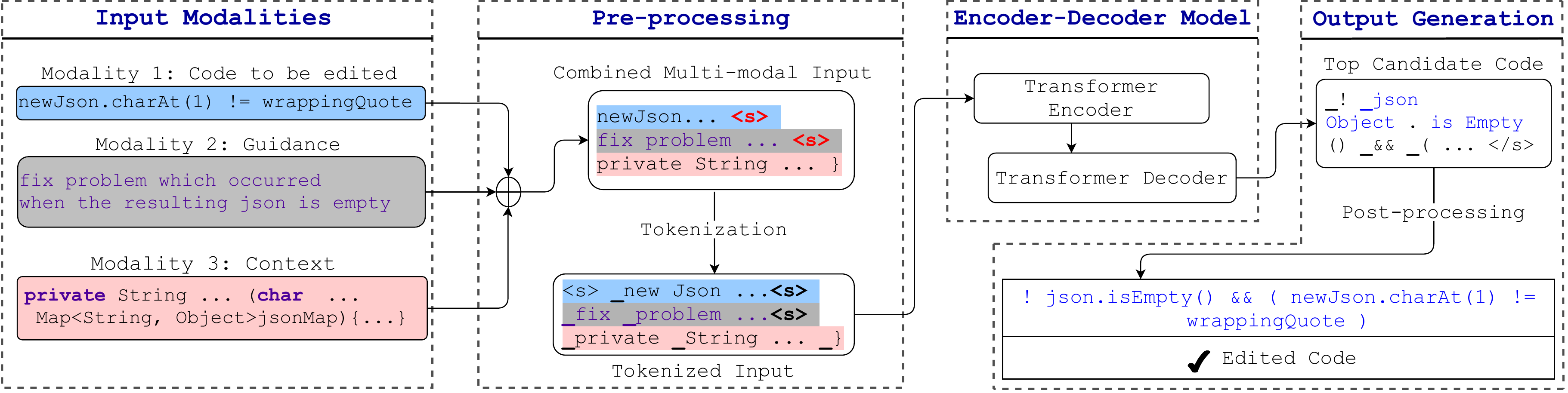}
    \vspace{1pt}
    \caption{\footnotesize \bf Overview of \tool pipeline}
    \label{fig:overview}
    
\end{figure*}
\Cref{fig:overview} shows an overview of \tool's working procedure. \tool is a multi-layer encoder-decoder based model consisting of a Transformer-based encoder and a Transformer-based decoder. Both the encoder and decoder consist of 6 layers. \tool works on three different modalities of information: (i) Code that needs to be edited (\buggy), (ii) natural language guidance from the developer (\guidance), and (iii) the context code where the patch is applied (\context).{
We acknowledge that \buggy is essentially a substring of \context. However, by explicitly extracting and presenting \buggy to \tool, we provide \tool with additional information about the change location. Thus, despite being a part of the context, we consider \buggy a separate modality.} Nevertheless, \tool consists of three steps. First, the pre-processing step processes and tokenizes these input modalities (\S\ref{tool/preprocessing}). Then the encoder in \tool encodes the processed input, and the decoder sequentially generates the patched code as a sequence of tokens (\S\ref{tool/modeling}). At final step, \tool post-processes the decoder generated output and prepares the edited code (\S\ref{tool/postprocess}).

\subsection{Pre-processing}
\label{tool/preprocessing}

\noindent \textit{Input Consolidation.} In the pre-processing step, \tool generates consolidated multi-modal input ($X$) from the three input modalities (\ie \buggy, \guidance, and \context). \tool combines these input modalities as a sequence separated by a special {\tt <s>} token \ie  \highlight{$X = $ \buggy~{\tt <s> }\guidance~{\tt <s> }\context}. In the example shown in \Cref{fig:motivating}, \buggy is \linecode{newJson.charAt(1)) != wrappingQuote}, \guidance is \texttt{fix problem which occurred when the resulting json is empty}, and \context is the whole function before the edit (see {\em Input Modalities} in \cref{fig:overview}). \tool generates a consolidates multi-modal input sequence as \texttt{newJson.charAt(1)) ... <s> fix problem which occurred ... <s> private String ... \}}. 

\smallskip

\noindent \textit{Tokenization.} \tool uses sentence-piece tokenizer~\cite{kudo-richardson-2018-sentencepiece}. Sentence-piece tokenizer divides every token into sequence of subtokens. Such subword tokenization is similar to previously used byte-pair encoding in automatic code editing literature~\cite{karampatsis2020big, jiang2021cure}. We use PLBART~\cite{ahmad2021unified}'s sentence-piece tokenizer which is trained on billions of code from GitHub. After tokenizing the consolidated input $X$ from \cref{fig:motivating}, we get \texttt{\U new Json . char At ( 1 ) ... <s> \U fix \U problem \U which \U oc cur red ... <s> \U private \U String ... \_\}}.


\subsection{Encoder-Decoder Model}
\label{tool/modeling}
The input to \tool's encoder-decoder model is a sequence of subtokens generated in the previous step. 

\smallskip

\noindent\textit{Transformer Encoder.} Given an input sequence $X = x_1, x_1, ..., x_n$, the encoder learns the representation of every token at layer $l$ as $R^e_l(x_i)$ using self-attention computed as
\begin{equation}
    R^e_l(x_i) = \sum_{j=i}^{n}{a_{i,j} * R^e_{l-1}(x_j)}
    \label{eqn:tool/encoder-representation}
\end{equation}
Where $R^e_{l-1}(x_j)$ is the representation of subtoken $x_j$ as generated by layer $l-1$, and $a_{i,j}$ is the attention weight of subtoken $x_i$ to $x_j$. Such attention weights are learned by multi-head attention~\cite{vaswani2017attention}. Final layer generated representation (\ie $R^e_6(x_i)$) is the final representation for every subtoken $x_i$ in the input. Note that, the encoder learns the representation of  \Cref{eqn:tool/encoder-representation} of a subtoken, using all subtokens in the sequence. Thus the learned representation of every subtoken contains information about the whole input sequence. Since we encode all the information modalities in one sequence, the learned representation of every subtoken encodes information about other modalities.

\smallskip

\noindent\textit{Transformer Decoder.} The decoder in \tool is a transformer-based sequential left-to-right decoder consisting of 6 layers. It sequentially generates one subtoken at a time using previously generated subtokens and the final representation (${R^e_{l}(x_i)}$) from the encoder. The decoder contains two modules -- (i) self-attention, and (ii) cross-attention. The self-attention layer work similar to the self-attention in the encoder. First, with self attention, decoder generates representation $R^d{l}(y_i)$ of last generated token $y_i$ with self attention on all previously generated tokens $(y_1, y_2, ..., y_i)$. This self attention follows same mechanism described in \Cref{eqn:tool/encoder-representation}. After learning decoder representation by self attention, decoder applies attention of encoder generated input representation using the following equation, 
\vspace{-2mm}
\begin{equation}
    \mathcal{D}_{l}(y_i) = \sum_{j=i}^{n}{\alpha^l_{i,j} * R^e_6(x_j)}
    \label{eqn:tool/decoder-representation}
\end{equation}
Where $
    \alpha^l_{i,j} = softmax\left(dot\left(R^e_6\left(x_j\right), R^d{l}\left(y_i\right)\right)\right)
$ is the attention weight between output subtoken $y_i$ to input subtoken $x_j$. The softmax generates an attention probability distribution over the length of input tokens. Finally the decoder learned representation, ${D}_{l}(y_i)$ is projected to the vocabulary to predict maximally likely subtoken from the vocabulary as next token.  

In summary, the encoder learns representation of every subtokens in the input using all input subtoken, essentially encoding the whole input information in every input subtoken representation. The decoder's self-attention mechanism allows the decoder to attend to all previously generated subtokens allowing the decoder decide on generating correct token at correct place. 
The cross-attention allows the decoder to attend to encoded representation - implicitly letting the model decide where to copy from the input where to choose from new tokens in the vocabulary. We initialize the end-to-end encoder-decoder in \tool using pre-trained weights of PLBART~\cite{ahmad2021unified}. 


\subsection{Output Generation}
\label{tool/postprocess}
The decoder in \tool continue predicting subtoken until it predicts the end of sequence {\tt </s>} token. During inference, \tool uses beam search to generate sequence of subtokens. Once the decoder finishes, \tool post-processes the top ranked sequence in the beam search. First, \tool removes the end of sequence {\tt </s>} token. It then detokenizes the subtokens sequence to code token sequence. In this step, \tool merges generated subtokens that are fragments of a code token into one code token. For the example shown in \cref{fig:motivating}, \tool generates the subtoken sequence \texttt{\U ! \U json . is Empty () \U\&\& \U( \U new Json . char At ( 1 ) \U!= \U wrap ping Quote \U) </s>}. After detokenization, \tool generates \linecode{! json.isEmpty()  && ( newJson.charAt(1) != wrappingQuote )}.

\section{Experimental Design}
\label{sec:method}

\subsection{Dataset}
\label{method/dataset}

\begin{table}[tbh]
    \centering
    \caption{\textbf{\footnotesize Statistics of the datasets studied.}}
    \label{tab:dataset-stat}
    \resizebox{\linewidth}{!}
    {
    \begin{tabular}{c|c|c|c|c|c|c}
    \hlineB{2}
    \multirow{2}{*}{\textbf{Dataset}} & {\textbf{Avg.}} & {\textbf{Avg.}} & {\textbf{Avg. tokens}} & \multicolumn{3}{c}{\textbf{\# examples}} \bigstrut[t]\\
    \cline{5-7}
    & \textbf{Tokens} & \textbf{Change Size*} & \textbf{in Guidance} & \textbf{Train} & \textbf{Valid} & \textbf{Test} \bigstrut[b]\\
    \hlineB{2}
    \sdata & 32.27 & 7.39 & 11.55 & 46628 & 5828 & 5831\bigstrut \\
    \hline
    \mdata & 74.65 & 8.83 & 11.48 & 53324 & 6542 & 6538\bigstrut \\
    \hlineB{2}
    \end{tabular}
    }
    {\footnotesize * Change size measured as token edit distance. }
    
\end{table}

To prove our concept of \tool, we experiment on two different datasets (\ie \sdata, and \mdata) proposed by Tufano~\etal~\cite{tufano2019empirical}. In these two datasets, they collected large collections of bug-fix code changes along with commit messages from Java projects in GitHub. Each example in these datasets contains the java method before the change (\pcode), the method after the change (\ncode), and the commit message for the change.  There are some examples ($< 100$) with corrupted bytes in the commit message, which we could not process. We excluded such examples from the dataset. \Cref{tab:dataset-stat} shows statistics of the two datasets we used in this paper. \sdata contains smaller methods with maximum token length 50, and \mdata contains bigger methods with up to 100 tokens in length. The average size of the change (edit distance) is 7.39, and 8.83 respectively, in \sdata and \mdata.

\subsection{Data Preparation}
\label{method/data-prep} 
For the datasets described in \cref{method/dataset}, we extract the input modalities and the expected output to train \tool. For every method pair (\ie before edit - \pcode, after edit - \ncode) in those dataset, we use GumTree~\cite{falleri2014fine} to extract a sequence of tree edit locations. We identify the root of the smallest subtree of \pcode's AST that encompasses all the edit operations. We call the code fragment corresponding to that subtree as {\em code to be edited}(\buggy) and used as \tool's first modality. Similarly, we extract the code corresponding to the smallest subtree encompassing all the edit operations from \ncode and use that as {\em code after edit}(\fixed). We use the commit message associated with the function pair as \tool's second modality, guidance(\guidance). Finally, we use the full method before edit (\pcode) as \tool's third modality, context(\context). 



\subsection{Training}
\label{method/training}

After combining every example in the datasets in \tool's input (\buggy, \guidance, \context) and expected output (\fixed), we use this combined dataset to train \tool. For training \tool, we use Label Smoothed Cross Entropy~\cite{muller2019does} as loss function. We use Adam optimizer, with a learning rate of $5e^{-5}$. We train \tool for 30 epochs, after every epoch, we run beam search inference on the validation dataset. We stop training if the validation performance does not improve for five consecutive validations. 

\subsection{Evaluation Metric}
\label{method/metric}
We use the top-1 accuracy as the evaluation metric throughout the paper. For proof-of-concept, we evaluate all techniques with beam size 5. When the generated patched code matches {\em exactly} with the expected patched code \fixed, it is correct, incorrect otherwise. Note that this is the most stringent metric for evaluation. Previous approaches~\cite{chakraborty2020codit, chen2019sequencer, lutellier2020coconut} talked about filtering out infeasible patches from a ranked list of top k patches using test cases. However, we conjecture that such test cases may not always be available for general purpose code edits. Thus, we only compare top-1 accuracy. 

\subsection{Research Questions}
\label{method/research_questions}

\tool contains several design components: (i) use of multimodal information, (ii) use of transformer and 
initializing it with the pre-trained model, and (iii) use of end-to-end encoder-decoder (using PLBART) to generate patches instead of separately using pre-trained encoder or pre-trained decoder, as used by previous tools. 
First, we are interested in evaluating \tool \wrt state-of-the-art methods. In particular, we evaluate how these three design choices effect \tool's performance. So, we start with investigating,  
\RQ{1}{\rqa}

\tool uses three input modalities. Our next evaluation target is how these individual modalities effect \tool's performance? Thus we ask, 
\RQ{2}{\rqb}

Finally, recall from \Cref{tool/preprocessing}, \tool proposes to encode all the input modalities as a sequence and use one encoder for the consolidated multi-modal input. An alternative to this encoding mechanism is to encode individual input modality with dedicated input encoder. Our next evaluation aims at finding out the best strategy to encode input modalities. Hence, we investigate, 
\RQ{3}{\rqc}

\section{Empirical Results}
\label{sec:results}

In our first research question, we evaluate \tool's performance \wrt other techniques and the effect of \tool's design components.

\RQrepeat{1}{\rqa}

\noindent{\bf Experimental Setup.}  We carefully chose the baselines to understand the contribution from different design choices of \tool. 
We evaluated our model in two experimental settings. First, we train different baseline models where the full model is trained from scratch. In this setting, the first baseline we consider is an LSTM with attention~\cite{bahdanau2014neural} NMT model. Various existing code patching approaches~\citep{tufano2018nmt_bug_fix, tufano2019empirical, tufano2019learning, chen2019sequencer} used such settings. Second baseline  is Transformer~\citep{vaswani2017attention} based \sts model. We consider two different-sized transformers. This enables us to contrast effect of model size in code-editing performance. The \transBase model consists of six encoder layers and six decoder layers. The \transBase model's architecture is the same as \tool's architecture. Furthermore, we consider another transformer with a much larger architecture. \transLarge contains twelve encoder layers and twelve decoder layers with three times as many learnable parameters as the \transBase model. The final baseline in this group is \codit, which is a tree-based model. Comparison \wrt  ~\codit allows us to contrast externally given syntax information (in the form of CFG) and learned syntax by transformers (\ie \tool). We use all three input modalities (see \Cref{fig:overview} for example) as input to the LSTM and Transformer. Using auxiliary modalities is non-trivial with \codit since the input to \codit must be a syntax-tree. Thus, we use uni-modal input (\buggy) with \codit.

\begin{figure}[tbh]
    \centering
    \begin{subfigure}{0.80\linewidth}
    \centering
    \includegraphics[width=0.95\linewidth]{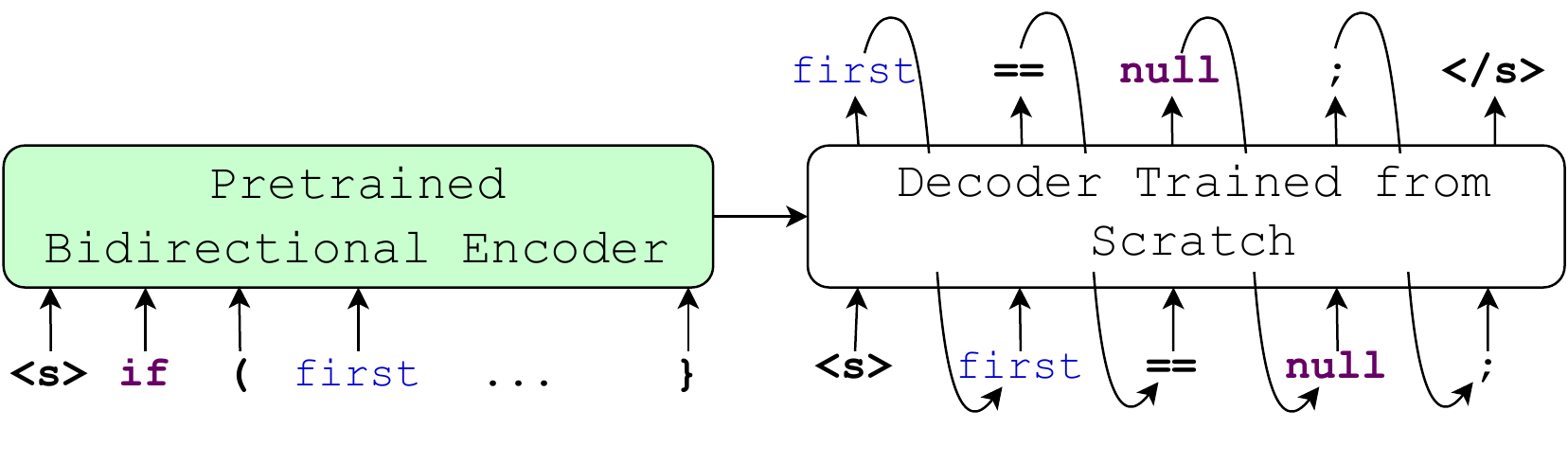}
    \vspace{-2mm}
    \caption{\textbf{\footnotesize CodeBERT \textemdash  ~Consist of bidirectional pretrained encoder and a decoder trained from scratch.}}
    \label{fig:codebert}
    \end{subfigure}
    
    \vspace{2mm}
    
    \begin{subfigure}{0.80\linewidth}
    \centering
    \includegraphics[width=0.95\linewidth]{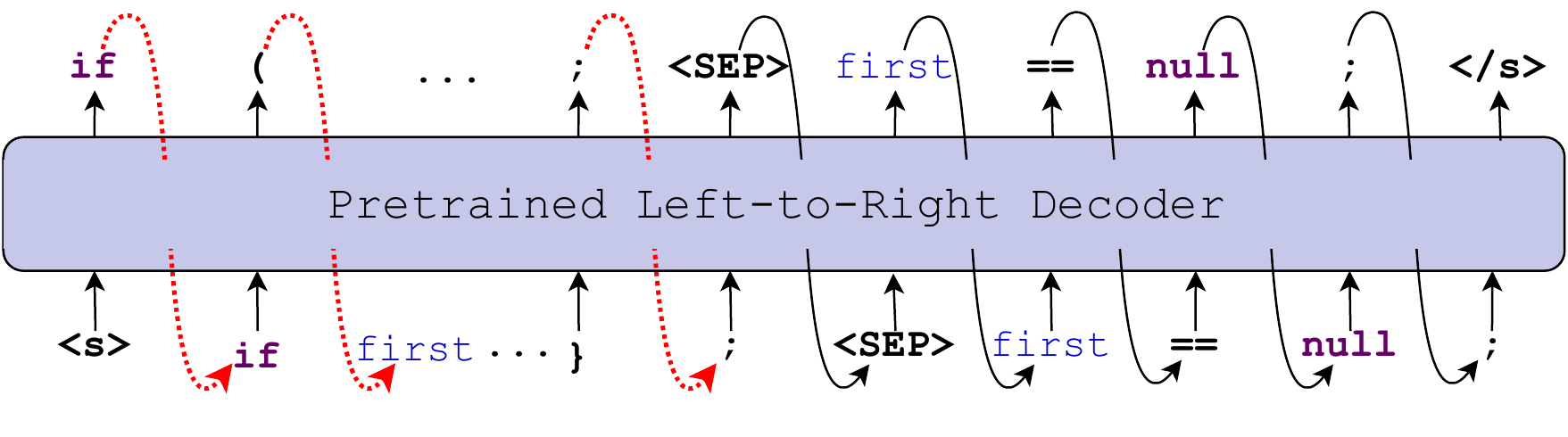}
    \vspace{-2mm}
    \caption{\textbf{\footnotesize CodeGPT \textemdash ~One pretrained single decoder processes the input and output sequentially from left to right.}}
    \label{fig:codeGPT}
    \end{subfigure}
    
    \vspace{2mm}
    
    \begin{subfigure}{0.80\linewidth}
    \centering
    \includegraphics[width=0.95\linewidth]{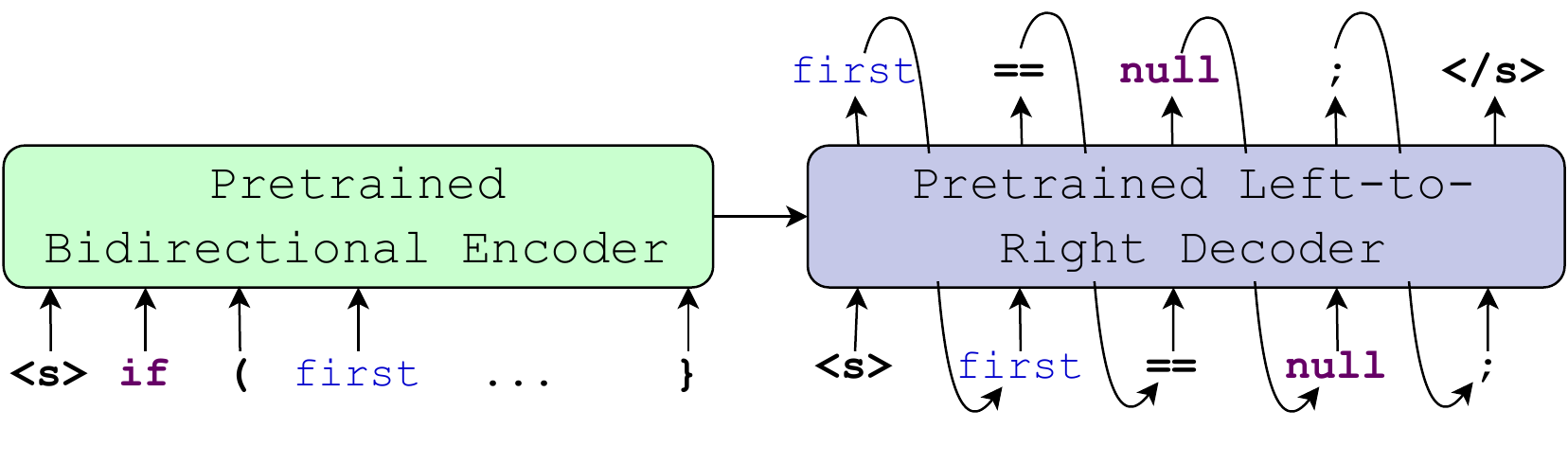}
    \vspace{-2mm}
    \caption{\textbf{\footnotesize PLBART \textemdash ~Consist of pretrained  bidirectional encoder and pretrained left to right decoder.} }
    \label{fig:plbart}
    \end{subfigure}
    
    \caption{\textbf{\footnotesize Schematic diagram of the three types of pre-trained models. used to evaluate \tool.} }
    \label{fig:three_models}
\end{figure}

In the second setting, we consider 
different pre-trained models, which we used to fine-tune for patch generation. \Cref{fig:three_models} shows schematic diagrams of the pre-trained models we compared in this evaluation. First two models we considered are CodeBERT~\citep{feng2020codebert}, and GraphCodeBERT~\citep{guo2020graphcodebert}. Both of these models are pretrained encoders primarily trained to understand code. To use these for the patching task, we add a six-layered transformer-based decoder along with the encoder. The decoder is trained from scratch (see \cref{fig:codebert}). Another pre-trained baseline is CodeGPT~\citep{CodeXGLUE}. GPT is a single left-to-right decoder model primarily pre-trained to generate code. For the code editing task, a special token {\tt <SEP>} combines the input and the output as a sequence separated. Jiang~\etal~\cite{jiang2021cure} showed the effectiveness of GPT for the source code patching task (see \cref{fig:codeGPT}). In contrast to these pre-trained models, \tool uses PLBART, an end-to-end encoder-decoder model trained to understand and generate code simultaneously (see \cref{fig:plbart}). To compare from a fairground, we evaluate these pre-trained models with uni-modal input (\buggy), and multi-modal input (\buggy {\tt <s>} \guidance {\tt <s>} \context), separately. 


\begin{table}[!tbh]
    \centering
    \scriptsize
    \caption{\footnotesize \textbf{Top-1 accuracies of different models \wrt their training type, model sizes, input modality.}}
    \label{tab:rq1-result}
    \resizebox{\linewidth}{!}
    {\begin{tabular}{c|l|c|c|c|c}
    \hlineB{2}
        \textbf{Training} & \textbf{Model} &  \textbf{\# of} & \textbf{Multi-} & \multicolumn{2}{c}{\textbf{Accuracy (\%)}} \bigstrut[t]\\
        \cline{5-6}
        \textbf{Type} & \textbf{Name} &  \textbf{params (M)} & \textbf{Modal} & \textbf{\sdata} & \textbf{\mdata} \bigstrut[b]\\
        \hlineB{2}
         & LSTM & 82.89 & \tick & 6.14 & 1.04  \bigstrut\\
        \cline{2-6}
         & \transBase & 139.22 & \tick & 11.18 & 6.61  \bigstrut\\
        \cline{2-6}
         & \transLarge & 406.03 & \tick & 13.40 & 8.63  \bigstrut\\
        \cline{2-6}
        \multirow{-5}{*}{\rotatebox{90}{~~From Scratch}}  & \sc{Codit} & 105.43 & \cross & 6.53 & 4.79  \bigstrut\\
        \hlineB{2}
        &  &  & \cross & 24.28 & 16.76  \bigstrut\\
        \cline{4-6}
         & \multirow{-2}{*}{CodeBERT} & \multirow{-2}{*}{172.50} & \tick & 26.05 & 17.13  \bigstrut\\
        \cline{2-6}
        &  &  & \cross & 24.44 & 16.85  \bigstrut\\
        \cline{4-6}
         & \multirow{-2}{*}{GraphCodeBERT} & \multirow{-2}{*}{172.50} & \tick  & 25.67 & 18.31  \bigstrut\\
        \cline{2-6}
        &  &  & \cross & 28.13 & 16.35  \bigstrut\\
        \cline{4-6}
         & \multirow{-2}{*}{CodeGPT} & \multirow{-2}{*}{124.44} & \tick & 28.43 & 17.64  \bigstrut\\
        \cline{2-6}
        &  &  & \cross & 26.67 & 19.79  \bigstrut\\
        \cline{4-6}
        \multirow{-10}{*}{\rotatebox{90}{Fine-tuned}} & \multirow{-2}{*}{\tool} & \multirow{-2}{*}{139.22} & \tick & \textbf{29.99} & \textbf{23.02}  \bigstrut\\
        \hlineB{2}
    \end{tabular}}
    \vspace{2mm}
    
\end{table}

\noindent{\bf Results.}~\Cref{tab:rq1-result} shows the accuracy in top 1 predicted patch by \tool along with different baselines. LSTM based \sts model predicted 6.14\% and 1.04\% correct patches in \sdata and \mdata respectively. The \transBase model achieves 11.18\% and 6.61\% top-1 accuracy in those datasets, which improves further to 13.40\% and 8.63\% with the \transLarge model. \codit predicts 6.53\% and 4.79\% correct patches in \sdata and \mdata, respectively. Note that \codit takes the external information in the form of CFG; thus, the patches \codit generate are syntactically correct. Nevertheless, the transformers, even the smaller model, perform better to predict the correct patch. We conjecture that the transformer model can implicitly learn the code syntax  without direct supervision.

In contrast to the models trained from scratch, when we fine-tune a pretrained model, it generates {\em significantly more} correct patches than models trained from scratch. For instance, \tool~{(initialized with pretrained PLBART)} generates {\em 168\%} and {\em 248\%} more correct patches than the \transBase model {(with randomly initialized parameters)}, despite both of these models having the same architecture and the same number of parameters. In fact, the smallest fine-tuned model (CodeGPT) performs much better than the larger model trained from scratch (\transLarge). 

All the fine-tuned models exhibit better performance when the input data are multi-modal with various degrees of improvement. With all three input modalities, CodeBERT~\cite{feng2020codebert} generates 7\% and 2.2\% more correct patches in \sdata and \mdata, respectively, compared to a unimodal CodeBERT model. In case of \tool, such improvement is 11.07\% in \sdata and 16.23\% in \mdata. The \guidance in the multi-modal data often contains explicit hints about how to change the code. For instance, consider the example shown in \Cref{fig:motivating}, the guidance explicitly says there is a problem with the {\tt json} when it is {\tt empty}. Furthermore, with the presence of \context in the input, the model can identify different variables, methods used in the method and potentially copy something from the context. We conjecture that such additional information from these two additional input modalities (i) reduce the search space for change patterns, (ii) help models copy relevant identifiers from the context. 

Among the fine-tuned models multi-modalities, \tool generates 15.12\% more correct patches than CodeBERT, 16.82\% than GraphCodeBERT, and 5.49\% than CodeGPT in \sdata. In the case of \mdata dataset, \tool's improvement in performance is 34.38\%, 25.72\%, 30.50\% higher than CodeBERT, GraphCodeBERT, and CodeGPT, respectively. To understand these results better, let us look at some of the examples. 

\begin{figure}[!tbh]
\small
\begin{subfigure}{0.92\linewidth}

\begin{tabular}{l}
\lstset{escapechar=~,style=CustomJava}
\begin{lstlisting}
~\textbf{//Guidance: merging of items that aren't actually equal}~
public static boolean equals(
               ItemStack one, ItemStack two) { 
~\red{\textbf{-}}~   ~\red{\textbf{return one.isSimilar(two) \&\&}}~
~\red{\textbf{-}}~           ~\red{\textbf{ (one.toString().equals(two.toString()));}}~
~\dkgreen{\textbf{+}}~   ~\dkgreen{\textbf{return one.isSimilar(two);}}~ //MODIT generated
    /* CodeGPT generated */
~\dkgreen{\textbf{+}}~   ~\blue{\underline{\textbf{return one.toString().equals(two.toString());}}}~ 
}
\end{lstlisting}
\end{tabular}
\end{subfigure}


\caption{\bf \footnotesize Example patch where \tool was able to generate correct patch, but CodeGPT could not. \tool's patch is shown in \dkgreen{green}, and CodeGPT generated patch is shown in \blue{\underline{blue}}.}
\label{fig:compare-with-gpt}
\end{figure}

\Cref{fig:compare-with-gpt} shows an example patch where \tool correctly generated the expected patch but CodeGPT could not. If we look closely, we can see that the code to be changed (\buggy) is a boolean expression where the two clauses are combines with {\tt \&\&}. While only the first clause, \linecode{one.isSimilar(two)} is the expected output, CodeGPT chooses the second clause, \linecode{one.toString().equals(two.toString())} from the original. Recall from \cref{fig:codeGPT}, CodeGPT processes the combined input and output sequence (separated by special {\tt <SEP>} token) in {\em left-to-right}  fashion. Thus, encodes representation of the input tokens do not contain information about the whole input sequence. In contrast, the \tool uses a pre-trained {\em bi-direction} encoder which helps \tool to understand the input fully. Based on the examples we have seen and the empirical result, we conjecture that, for code-editing tasks, the model must fully understand the input in a bi-directional fashion. 

\begin{figure}[!tbh]
\small
\begin{subfigure}{0.92\linewidth}
\begin{tabular}{l}
\lstset{escapechar=~,style=CustomJava}
\begin{lstlisting}
~\textbf{// Guidance: ... code refactoring ...}~
public boolean isEmpty() { 
~\red{\textbf{-}}~  ~\red{\textbf{if((first) == null)\{ return true;\}}}~ 
~\red{\textbf{-}}~  ~\red{\textbf{return false;}}~
~\dkgreen{\textbf{+}}~  ~\dkgreen{\textbf{return (first) == null;}}~ //MODIT predicts
   /* CodeBERT generated */
~\dkgreen{\textbf{+}}~  ~\blue{\textbf{\underline{return ((first) == null) || (first.get()) == null;}}}~ 
}
\end{lstlisting}
\end{tabular}
\end{subfigure}


\caption{\textbf{\footnotesize Correctly predicted patch by \tool. CodeBERT could not understand and reason about the textual hint to predict the correct patch.}}
\label{fig:compare-with-bert}
\end{figure}

\Cref{fig:compare-with-bert} shows an example where \tool generated correct patch, CodeBERT could not. Note that the guidance text explicitly asks about {\em code refactoring}, implying that the patched code should be semantically similar to the original code. Similar to the original code, patched could should return \linecode{true} when \linecode{first == null}, otherwise it should return \linecode{false}. An automated code change tool should not add additional code features when doing the refactoring. 
However, CodeBERT generated patch which introduced an additional clause \linecode{first.get() == null} in the return expression, which make CodeBERT's generate code semantically different from the original. \tool was able to generate the correct patch for this example. 

Finally, we summarize the empirical lessons we learned in this research question as

\begin{itemize}
    \item Multi-modal input improves Code-Editing capability, irrespective of the underlying model used. The guidance often narrows the edit pattern search space, and the context narrows down the token generation search space.
    \item Transformer models (especially larger ones) are robust enough to learn the code's syntax information without direct supervision. When a pre-trained model is used to initialize transformer parameters, the improvement is {\em notably higher}.
    \item For code-editing task, both {\em understanding the input} and {\em correctly generated} output are important. While a pre-trained encoder understands the code and a pre-trained decoder generates correct code, an end-to-end pre-trained encoder-decoder model (\eg PLBART) the best choice to fine-tune for this task.
\end{itemize}

\vspace{1mm}

\RS{1}{\tool generates 29.99\%, and 23.02\% correct patches in top-1 position for two different datasets outperforming CodeBERT by up to 25.72\%, GraphCodeBERT by up to 34.38\%, and CodeGPT by up to 30.50\%. {Pre-trained models tend to be more effective than models trained from scratch for code editing---\tool improves the performance by 167\% than the best model trained from the scratch.} }



\Comment{
\smallskip 

\noindent{\bf Few case studies.}
\begin{figure}[!tbh]
\small

\begin{subfigure}{0.92\linewidth}
\caption{\textbf{\small Expected Patch.}}
\vspace{-2mm}
\label{fig:bigger-mistake-expected}
\begin{tabular}{l}
\lstset{escapechar=~,style=CustomJava}
\begin{lstlisting}
~\textbf{// Guidance : <s> fixed prices api bug}~ 
public static Map<String, Object>priceMap (
                                Double price) { 
    return new HashMap<String, Object>(){{
~\dkgreen{\textbf{+}}~       ~\dkgreen{\textbf{put("id", 1);}}~
        put("price", price);
    }}; 
}
\end{lstlisting}
\end{tabular}
\end{subfigure}
\vspace{2mm}

\begin{subfigure}{0.92\linewidth}
\caption{\textbf{\footnotesize \tool predicted patch.}}
\vspace{-2mm}
\label{fig:bigger-mistake-tool}
\begin{tabular}{l}
\lstset{escapechar=~,style=CustomJava}
\begin{lstlisting}
public static Map<String, Object>priceMap (
                                Double price) { 
    return new HashMap<String, Object>(){{
~\dkgreen{\textbf{+}}~       ~\dkgreen{if(price != null)}~
            put("price", price);
    }}; 
}
\end{lstlisting}
\end{tabular}
\end{subfigure}
\vspace{2mm}

\begin{subfigure}{0.92\linewidth}
\caption{\textbf{\footnotesize CodeGPT predicted patch.}}
\vspace{-2mm}
\label{fig:bigger-mistake-gpt}
\begin{tabular}{l}
\lstset{escapechar=~,style=CustomJava}
\begin{lstlisting}
public static Map<String, Object>priceMap (
                                Double price) { 
    return new HashMap<String, Object>(){{
~\dkgreen{\textbf{+}}~       ~\dkgreen{if(price < 0) \{ price = 0;\}}~
        put("price", price);
    }}; 
}
\end{lstlisting}
\end{tabular}
\end{subfigure}
\vspace{2mm}

\begin{subfigure}{0.92\linewidth}
\caption{\textbf{\footnotesize CodeBERT predicted patch.}}
\vspace{-2mm}
\label{fig:}
\label{fig:bigger-mistake-codebert}
\begin{tabular}{l}
\lstset{escapechar=~,style=CustomJava}
\begin{lstlisting}
public static Map<String, Object>priceMap (
                                Double price) { 
    return new HashMap<String, Object>(){{
~\red{\textbf{-}}~       ~\red{\textbf{put("price", price);}}~
~\dkgreen{\textbf{+}}~       ~\dkgreen{\textbf{priceMap(price);}}~
~\dkgreen{\textbf{+}}~       ~\dkgreen{\textbf{return price;}}~
    }}; 
}
\end{lstlisting}
\end{tabular}
\end{subfigure}
\caption{\textbf{\footnotesize Example where none of the pretrained model succeeded to generate the correct patch. Both CodeBERT and CodeGPT generated a patch which break the semantics of the code. \tool's predicted patch maintained semantic correctness in the code.}}
\label{fig:bigger-mistake-by-bert}
\end{figure}
Let us look at another example shown in \Cref{fig:bigger-mistake-by-bert}. \Cref{fig:bigger-mistake-expected} shows the expected patch and corresponding textual guidance given to the model. A Closer look into the guidance and expected patch reveals that it is a tough patch to generate, even for a human developer, unless the developer has a profound knowledge about the issue. Thus, it is highly challenging, if not impossible, for an automated tool to generate the correct patch in such a case. As expected, none of the models we considered generated the correct patch. However, to better understand the models, let us look at each model's top patches. 

\tool added an \linecode{if} block surrounding the existing \linecode{put("price", price);} statement (see \Cref{fig:bigger-mistake-tool}). While this is not the expected patch, the patched code is still compilable and syntactically, and semantically correct. CodeGPT (see \Cref{fig:bigger-mistake-gpt}) produces a patch by adding a complete if block where it sets the value of \linecode{price} to \linecode{0} if existing value is less than \linecode{0}. While the patched code by CodeGPT is syntactically correct, semantically, it is not. Because the {\em local variable \linecode{price} defined in an enclosing scope must be final or effectively final.}\footnote{Error message as shown by Eclipse IDE.} CodeBERT predicts a patch by removing the existing \linecode{put("price", price);} and adding a function call statement to the function itself with the same parameter, effectively making the potential for infinite recursion. After the statement CodeBERT added a {\tt return} statement. However, control of the code cannot return from an {\em object initializer scope}. Thus, the initializer scope should not contain any return statement.

\rayb{You made a big deal out of syntactic and semantic correct code. I think that is PLBART's contribution not \tool.}

\noindent{\bf Explanation of the results.} To understand such performance differences between different pretrained models, let us analyze how we used those models for the code change modeling task.  \Cref{fig:three_models} shows conceptual schematic diagrams for different pretrained models. CodeBERT (also GraphCodeBERT) pretrains a bidirectional encoder. From a conceptual point of view, any BERT~\cite{devlin2018bert} model learns to understand the input. To use the BERT model in generative tasks such as code editing, the generator(decoder) is trained from scratch~\cite{feng2020codebert}. Thus, the decoder used with CodeBERT may not have robust knowledge about generating {\em syntactically and semantically correct} code. 

CodeGPT uses one left-to-right (autoregressive) decoder and uses that for pretraining. Such decoder is pretrained on large corpora of code. The pretraining objective of CodeGPT is to learn how to generate syntactically and semantically correct code. Since CodeGPT is a single decoder, to use it in code change modeling task, the input is followed by the output with a special separator token ({\tt <SEP>}) in between~\cite{jiang2021cure}. During inference, the input tokens are processed by the CodeGPT from left to right till the {\tt <SEP>} token; after that, CodeGPT starts generating code with beam search. While performing the beam search, CodeGPT learns the attention weights across the input token representations generated earlier. Since CodeGPT generates the representation of input tokens in a unidirectional way only, such representations are less robust than those generated by CodeBERT. 

\tool combines the advantages of both CodeBERT and CodeGPT. \tool consists of a bidirectional encoder and an autoregressive decoder. \tool pre-trains both the encoder and decoder in an end-to-end fashion using denoising autoencoding~\cite{lewis2019bart}. \tool's encoder learns to generate robust input representation,  while the decoder learns to generate syntactically and semantically correct code. We attribute \tool's improved performance over both CodeBERT and CodeGPT to pre-trained robust bidirectional encoder and autoregressive decoder. 
}

\tool combines multiple modalities of information to generate patches. Now we investigate, 

\RQrepeat{2}{\rqb}

\noindent{\bf Experimental Setup.} In this experiment, we investigate the contribution of different input modalities in \tool's performance. Recall from ~\cref{tool/preprocessing} that we use three inputs in \tool (\ie \buggy, \context, \guidance). Here, we investigate different combinations of such input modalities. 
{More precisely, we investigate the influence of three information sources: (i) code that needs to be changed (\buggy), (ii) context (\context), and (iii) guidance (\guidance). 
Note that, by presenting \buggy as a separate information modality, we are essentially providing \tool with the information about the location of the change. To study the effect of such presentation, we study another alternative experimental setup, where we annotate the change location inside the context with two unique tokens {\tt<START>} and {\tt <END>}.
}

\begin{table}[!tbh]
    \centering
    \scriptsize
    \caption{\textbf{\footnotesize Contribution of different input modalities in \tool's performance. \tick ~
    indicates that corresponding input modality is used as encoder input, \cross ~indicates otherwise. 
    We report top-1 accuracy as performance measure. Exp. ID is used later to refer to corresponding experiment result. Exp. ID $\Phi_{*}$ denotes an experiment with $*$ as input modalities.}}
    {\begin{tabular}{c|c|c|c|c|c}
    \hlineB{2}
    \multirow{3}{*}{\textbf{Exp. ID}} &\multicolumn{3}{c|}{\textbf{Inputs}} &  \multicolumn{2}{c}{\textbf{Accuracy (\%)}} \bigstrut\\
    \cline{2-6}
    & \multirow{1}{*}{\textbf{\buggy}} & \multirow{1}{*}{\textbf{\context}} & \multirow{1}{*}{\textbf{\guidance}} & \textbf{\sdata} & \textbf{\mdata} \bigstrut\\
    \hlineB{2}
    $\Phi_{c}$ & \cross & \tick & \cross & 13.05 & 4.50 \bigstrut \\
    \hline
    $\Phi_{cg}$ & \cross & \tick & \tick & \textbf{17.89} & \textbf{4.51} \bigstrut \\
    \hlineB{2}
    $\Phi_{c}^\dagger$ & \cross & \tick$^\dagger$ & \cross & 13.03 & 4.53 \bigstrut \\
    \hline
    $\Phi_{cg}^\dagger$& \cross & \tick$^\dagger$ & \tick & \textbf{17.90} & \textbf{4.60} \bigstrut \\
    \hlineB{2}
    $\Phi_{e}$ & \tick & \cross & \cross & 26.67 & 19.79 \bigstrut \\
    \hline
    $\Phi_{eg}$ & \tick & \cross & \tick & \textbf{28.76} & \textbf{21.63} \bigstrut \\
    \hline
    $\Phi_{ec}$ & \tick & \tick & \cross & 29.79 & 21.40 \bigstrut \\
    \hline
    $\Phi_{ecg}$ & \tick & \tick & \tick & \textbf{29.99} & \textbf{23.02} \bigstrut \\
    \hlineB{2}
    \end{tabular}
    }\\
    $^\dagger$ \buggy is surrounded by two special tokens \texttt{\textbf{<START>}} and \texttt{\textbf{<END>}} inside the context.

    \label{tab:modality_comparison}
\end{table}

\noindent{\bf Result.} \Cref{tab:modality_comparison} shows \tool's performance with different combination of input modalities. When we present only the context to \tool, it predicts 13.05\% correct patches in \sdata and 4.50\% in the \mdata, which improves further to 17.89\%, and 4.51\% in those two datasets respectively when we add \guidance. Note that in these two scenarios, the model does not explicitly know which portion of the code needs to be edited; it sees the whole method and predicts (only) the patched code (\fixed). In addition to learning how to patch, the model implicitly learns where to apply the patch in this setup. To test whether the identification of such location is the performance bottleneck, we surround the code that needs to be patched with two special tokens \texttt{<START>} and \texttt{<END>}.  SequenceR~\cite{chen2019sequencer} also proposed such annotation of buggy code. Surprisingly, such annotation resulted in comparable (slightly worse in one case) performance by \tool. 

In the next set of experiments, we extract the code that needs to be edited (\buggy) and present it as a separate input modality. First, we only present the \buggy without the other two modalities. When we only present the \buggy and generate the edited code (\fixed), it results in 26.67\% top-1 accuracy in the \sdata and 19.79\% in the \mdata. Ding~\etal~\cite{ding2020patching} attributed such improvement to the reduced search space due to shorter input. Our result corroborates their empirical findings. Nevertheless, when we add the \guidance modality with the \buggy, \tool's performance improves to 28.76\% and 21.63\% in \sdata and \mdata, respectively. 

In our final set of experiments in this research question, we augment \buggy with the \context. In this evaluation setup, \tool predicts 29.79\% correct patches in the \sdata and 21.40\% in the \mdata, which is improved further to 29.99\%, and 23.02\% correct patches in those two datasets when we add \guidance. 

\begin{figure}[!tbh]
\small

\begin{subfigure}{0.94\linewidth}
\begin{tabular}{l}
\lstset{escapechar=~,style=CustomJava}
\begin{lstlisting}
~// \textbf{Guidance:} fixed some bugs \dkgreen{\textbf{in type checking}}  ~
~// improved performance by caching types of \dkgreen{\textbf{expressions}}~
private TypeCheckInfo getType(SadlUnionType expression){ 
   ...
   return new TypeCheckInfo(
~\red{\textbf{-}}~       ~\red{\textbf{declarationConceptName, declarationConceptName}}~
        /* MODIT generated patch with guidance */
~\dkgreen{\textbf{+}}~       ~\dkgreen{\textbf{declarationConceptName, declarationConceptName, }}~
~\dkgreen{\textbf{+}}~       ~\dkgreen{\textbf{this, expression}}~
        /* MODIT generated patch without guidance */
~\dkgreen{\textbf{+}}~       ~\blue{\textbf{\underline{this.declarationConceptName,}}}~
~\dkgreen{\textbf{+}}~       ~\blue{\textbf{\underline{this.declarationConceptName}}}~
   ); 
} 
\end{lstlisting}
\end{tabular}
\end{subfigure}


\caption{\textbf{\footnotesize Example showing the effect of textual guidance in \tool's performance. \tool produced the \dkgreen{\bf correct patch} with guidance, without guidance as input \tool's produced \blue{\underline{\bf patch}} is essentially refactored version of original input.}}
\label{fig:effect-of-guidance}
\end{figure}
\Cref{fig:effect-of-guidance} shows an example where \tool with all modalities could successfully generate correct patch.
The text guidance (\guidance) provides hint that variable \linecode{expression} should somehow associate with the construction of  \linecode{TypeCheckInfo} in the \dkgreen{\bf patched code}. However, without this guidance \tool generated a \blue{\textbf{\underline{wrong patch}}} by accessing existing parameters from \linecode{this} object. Essentially, without the guidance, \tool refactored the input code. 

\begin{figure}[!tbh]
\small
\begin{subfigure}{0.92\linewidth}
\begin{tabular}{l}
\lstset{escapechar=~,style=CustomJava}
\begin{lstlisting}
~\textbf{// Guidance:} Fix bug of \textbf{\dkgreen{sending wrong message}}~
public void setPredecessor (model.Message m) { 
    this.predecessor = Integer.valueOf(m.Content);
    model.Message ~{\textbf{sent}}~ = new model.Message(); 
    sent.To = m.Origin; 
~\red{\textbf{-}}~   ~\red{\textbf{sendMessage(m)}}~; 
    /* MODIT generates with the context. */
~\dkgreen{\textbf{+}}~   ~\dkgreen{\textbf{sendMessage(sent)}}~; 
    /* MODIT generates without context as input. */
~\dkgreen{\textbf{+}}~   ~\blue{\underline{\textbf{sendMessage(m.toString())}}}~; 
}
\end{lstlisting}
\end{tabular}
\end{subfigure}


\caption{\textbf{\footnotesize Example showing the necessity of context information in predicting the correct patch. \tool's generated \dkgreen{correct patch} with the context as input. Without context, \tool received {\tt sendMessage(m)} and the guidance as input, did not know the variable {\tt sent} could be the parameter of the function {\tt sendMessage}, and predicted a \blue{\underline{wrong patch}}.}}
\label{fig:effect-of-context}
\end{figure}
\Cref{fig:effect-of-context} shows the effect of context as input modality to \tool. The before edit version of the code(\buggy) passed the wrong parameter (\linecodewospace{m}) to \linecode{sendMessage} function. When the context (\context) is presented to \tool, it saw another variable (\linecodewospace{sent}) in the context. In contrast, without context(\context), \tool indeed changed the parameter; but sent \linecodewospace{m.toString()} --- resulting in a wrong patch. 


When we extract the buggy code and present the buggy code along with the context, we see a big performance improvement (see the difference between $\Phi_{c}$, and $\Phi_{ec}$ in \cref{tab:modality_comparison}). 
We hypothesize that, when only context (\ie full code) is presented ($\Phi_{c}$), the model gets confused to identify which portion from the context needs to be edited since any portion of the code is a likely candidate for patching. However, when we extract the exact code that needs to be edited and present as a separate input modality to \tool, it can focus on patching just that code using other modalities (including the context) as a supporting source of information. 
In a recent study, Ding~\etal~\cite{ding2020patching} pointed out the need for effective ways to include context in the NMT based code editors. Our empirical results show that \tool's way of including context as a separate modality is a potential solution to that problem.

{In summary, each  of  the  modalities contribute to the overall performances of \tool.}  Lessons learned in these experiments are: 
\begin{itemize}
    \item {
    Additional textual guidance helps the patch generation. Such guidance can provide important clue about how to modify the code and sometimes provide ingredients necessary for the change.}
    \item Adding context explicitly in the input enables the model to select appropriate identifiers for patching.
    \item Isolating buggy code help the model put proper focus on the necessary part of the code while leveraging auxiliary information from other modalities. 
\end{itemize}

\RS{2}{All three modalities (code to be edited, context, and guidance) are essential for \tool to perform the best. Without either one of those, performance decreases. \tool's performance improves up to 37.37\% when additional textual guidance is used as an input modality. Context modality improves \tool's performance up to 6.4\%.}

We investigate alternative ways to combine multiple input modalities. We ask, 
\RQrepeat{3}{\rqc}

\begin{figure}
    \centering
    \includegraphics[width=0.80\linewidth]{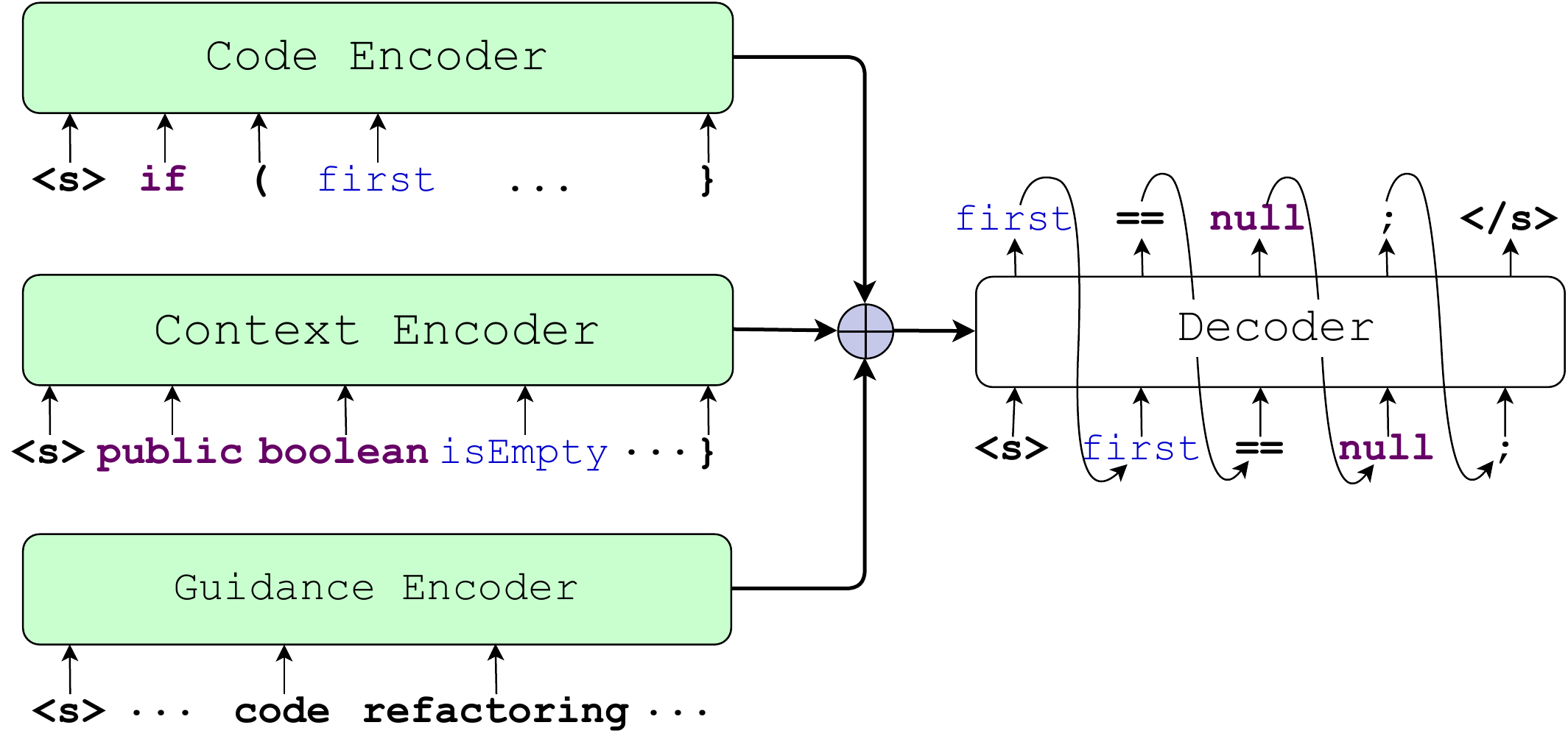}
    \caption{\textbf{\footnotesize An alternative architecture of code editing with multi-encoder model. We initialize each of the encoders with pre-trained Encoder model.}}
    \label{fig:multi-encoder}
\end{figure}

\noindent{\bf Experimental setup.} To validate \tool's design choice of appending all input modalities into one sequence, we test alternative ways to combine input modalities. In particular, we follow the design choice proposed by Lutellier~\etal~\cite{lutellier2020coconut}, where they used multiple encoders to encode the \buggy and the \context. Tufano~\etal~\cite{tufano2021towards} also leverages a similar idea to encode input code and code review messages. 
Nevertheless, we use a multi-encoder model shown in \cref{fig:multi-encoder}. 
In a multi-encoder setting, we first encode each input modality with a corresponding dedicated encoder. After the encoder finishes encoding, we concatenate the encoded representations and pass those to the decoder for generating patched code. To retain maximum effectiveness, we initialize each individual encoder with pretrained weights from CodeBERT~\cite{feng2020codebert}. We consider a single-encoder model (also initialized with CodeBERT) as a baseline to compare on the fairground. While presenting the inputs to the single encoder model, we concatenate input modalities with a unique separator token {\tt <s>}. Finally, to test the robustness of our empirical finding, we propose two different experimental settings. In the first evaluation setup, we use all three input modalities. We compare a tri-encoder model with a single-encoder model. Next, we consider bimodal input data -- \buggy and \guidance. We use a dual-encoder model and compare it with a single-encoder model in this setup.

\smallskip


\begin{table}[!tbh]
    \centering
    \scriptsize
    \caption{\textbf{\footnotesize Comparison of multi encoder model.}}
    \label{tab:multi-encoder-comparison}
    \begin{tabular}{c|c|c|c}
         \hlineB{2}
         \textbf{\# of} & \textbf{\# of}  & \multicolumn{2}{c}{\textbf{Accuracy (\%)}} \bigstrut[t]\\
         \cline{3-4}
         \textbf{Modalities} & \textbf{Encoders} & \textbf{\sdata} & \textbf{\mdata} \bigstrut[b]\\
         \hlineB{2}
         \multirow{2}{*}{3 (\buggy, \guidance, \context)} & 3 & 20.63 & 11.69 \bigstrut\\
         \cline{2-4}
         & \textbf{1} & \textbf{26.05} & \textbf{17.13} \bigstrut \\
         \hlineB{2}
         \multirow{2}{*}{2 (\buggy, \guidance)} & 2 & 23.12 & 15.49 \bigstrut\\
         \cline{2-4}
         & \textbf{1} & \textbf{23.81} & \textbf{17.46} \bigstrut \\
         \hlineB{2}
    \end{tabular}
    \vspace{2mm}
    
\end{table}
\noindent{\bf Result.} \Cref{tab:multi-encoder-comparison} shows the result of multi-encoder models. For tri-modal input data, if we use three different encoders, the model can predict 20.63\% correct patches in the \sdata and 11.69\% correct patches in the \mdata. In contrast, if we use a single encoder, the model's predictive performance increases to 26.05\% and 17.13\% top-1 accuracy in the \sdata and the \mdata, respectively. 

In the bimodal dataset (where the input modalities are \buggy and \guidance), the dual-encoder model predicts 23.12\% correct patches in the top-1 position for the \sdata and 15.49\% correct for the \mdata. The single encoder counterpart, in this setup, predicts 23.81\% correct patches for the \sdata and 17.46\% for the \mdata. The empirical results show that the single-encoder model performs better in both the experimental setup than the multi-encoder setup. We find similar results with GraphCodeBERT~\cite{guo2020graphcodebert}. 

\begin{figure}[tbh]
    \centering
    
    \begin{subfigure}{0.95\linewidth}
    \centering
    \includegraphics[width=0.65\linewidth]{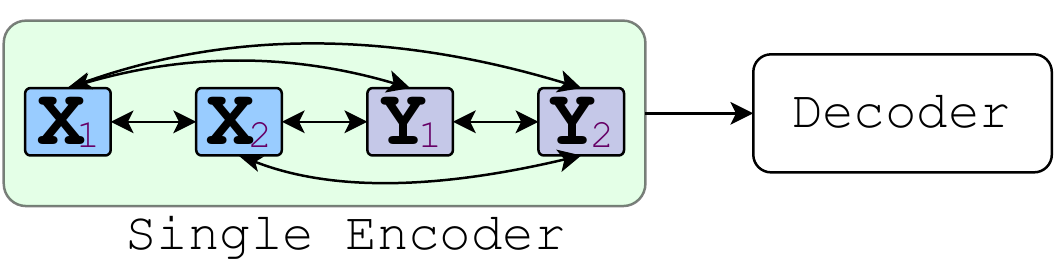}
    \caption{\textbf{\footnotesize 
    Single encoder for encoding multiple-modalities. Encoder can learn representation \wrt all modalities.
    }}
    \label{fig:single-encoder-explanation}
    \end{subfigure}
    
    \vspace{1mm}
    
    \begin{subfigure}{0.95\linewidth}
    \centering
    \includegraphics[width=0.65\linewidth]{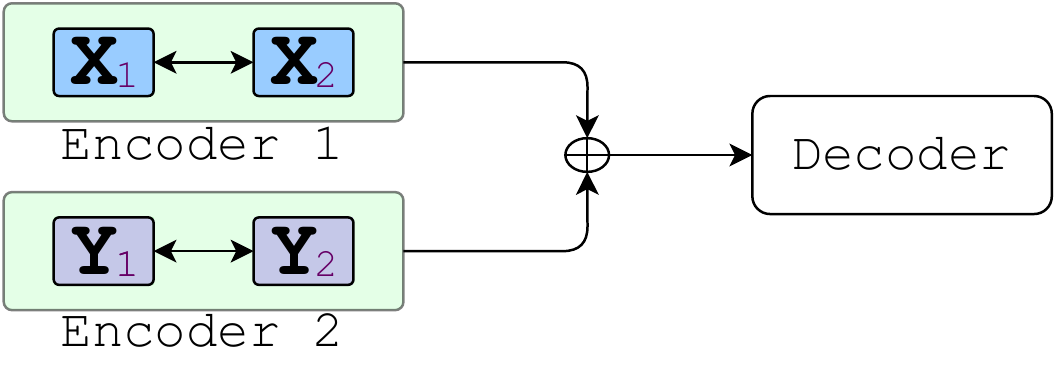}
    
    \caption{\textbf{\footnotesize Dual-encoder for encoding individual modalities separately. Representation of tokens from a particular modality is learned \wrt (only) other tokens from the same modality.}}
    \label{fig:dual-encoder-explain}
    \end{subfigure}
    
    \caption{\textbf{\footnotesize Input token representation generation in single encoder and multiple encoder.}}
    \label{fig:multi-encoder-explanation}
    \vspace{-2mm}
\end{figure}

\smallskip

To explain why single-encoder is performing better than multi-encoder, let us look at the encoders' working procedure. \Cref{fig:multi-encoder-explanation} depicts how the encoder generates representation for input tokens. Note that the encoders we used in this research question are transformer-based, and recall from the \cref{sec:background}, transformer generates representation for an input token by learning its dependency on all other tokens in the sequence. When we present all the input modalities to a single encoder, it generates input representation for those tokens \wrt and other tokens in the same modality and tokens from other modalities. For instance, in \cref{fig:single-encoder-explanation}, the encoder generates $X_2$'s representation considering $X_1$, $Y_1$, and $Y_2$. In contrast, in \cref{fig:multi-encoder-explanation}, $X_2$'s representation is learned only \wrt $X_1$, since encoder1 does not see the input modality $Y$. Thus, when we present all the input modalities to one single encoder, we conjecture that learned representations are more robust than that of learning with multi-encoder. 

Finally, we summarize the lessons we learned in this research question as

\begin{itemize}
    \item In multi-modal translation, using single encoder results in better performance than using a separate encoder for each modality. 
    \item Single-encoder generates input representation by inter-modality reasoning (attention), hence learns more robust representation than that of multi-encoder. 
\end{itemize}

\RS{3}{Encoding all the input modalities by a single encoder is the best way to learn in a multi-modal setting. A single encoder improves code-editing performance by up to 46.5\% than the corresponding multi-encoder setting.}



\section{Discussion}
\label{sec:discussion}

\subsection{Localization of Code Edit Site}
\label{subsec:discussion/localization}

An alternative modeling approach for code editing is to generate the sequence of edit operations (\ie {\tt ~INSERT, DELETE, UPDATE})~\cite{ding2020patching, dinella2019hoppity, tarlow2020learning, yao2021learning}, where the model must know the precise location of an edit operation (often a node in the AST) before applying it. Throughout this paper, we also assumed that such edit location is known to \tool. This assumption may pose a threat to the usefulness of \tool in a real development scenario. To mitigate such a threat, we perform an experiment where we pass the whole function as input to \tool and expect the whole edited function to be generated. 
\begin{table}[!tbh]
    \centering
    \scriptsize
    \caption{\textbf{\footnotesize Performance of \tool when the input in the full code and the output is patched full code.}}
    \label{tab:localization-exp}
    \begin{tabular}{c|c|c|c}
         \hlineB{2} 
         \multicolumn{2}{c|}{\textbf{Inputs}} & \multicolumn{2}{c}{\textbf{Accuracy (\%)}} \bigstrut \\
         \hline
         \textbf{Full Code} & \textbf{Guidance} & \textbf{\sdata} & \textbf{\mdata} \bigstrut\\
         \hlineB{2}
         \tick & \cross & 20.35 & 8.35 \bigstrut\\
         \hline
         \tick & \tick & \textbf{21.57} & \textbf{13.18} \bigstrut \\
         \hlineB{2}
    \end{tabular}
    \vspace{2mm}
    
\end{table}
\Cref{tab:localization-exp} shows the top-1 accuracy in the \sdata and the \mdata. \tool generates correctly patched full code in 20.35\% cases for the \sdata and 8.35\% cases for the \mdata. With additional textual guidance, the performance is further improved to 21.57\% and 13.18\% in the \sdata and \mdata, respectively. While textual guidance helps in this experimental setup, we notice a big drop in performance than the results shown in \cref{tab:modality_comparison}. This is because the benchmark datasets we used contain small edits (see \cref{tab:dataset-stat}). Thus, while generating the full code, the model wastes a large amount of effort trying to generate things that did not change. Nevertheless, our hypothesis {\em external guidance improves code editing} holds even when the model generates full code. 

\subsection{Tokenization for Source Code Processing}
\label{subsec:discussion/tokenization}

\begin{table}[h]
    \centering
    \scriptsize
    \caption{\textbf{\footnotesize Comparison between concrete tokenization and abstract tokenization alongside pre-trained models. Results are shown as top-1 accuracy of full code generation in \sdata / \mdata datasets.}}
    \label{tab:tokenization-discussion}
    \begin{tabular}{c|c|c|c}
    \hlineB{2}
    \textbf{Token type} & \textbf{CodeBERT} & \textbf{GraphCodeBERT} & \textbf{PLBART} \bigstrut\\
    \hlineB{2}
    Abstract & 16.4 / 5.16 & \textbf{17.30} / \textbf{9.10} & 19.21 / \textbf{8.98}\bigstrut\\
    \hline
    Concrete & \textbf{17.3} / \textbf{8.38} & 16.65 / 8.64 & \textbf{20.35} / 8.35 \bigstrut\\
    \hlineB{2}
    \end{tabular}
\end{table}

The possible number of source code can be virtually infinite. Vocabulary explosion has been a big challenge while processing source code with Machine Learning technique~\cite{tufano2018nmt_bug_fix, RMKarampatsis2019SStuB}. Previous research efforts have addressed this problem using several different heuristics. For instance, Tufano~\etal~\cite{tufano2018nmt_bug_fix, tufano2019empirical}  
identifiers abstraction, which drastically reduces the vocabulary size considered making it easier to learn patterns by the model. 
Recent studies~\cite{ding2020patching, lutellier2020coconut, jiang2021cure, RMKarampatsis2019SStuB} found that Byte-Pair Encoding~\cite{sennrich2015neural} partially solves the open-vocabulary problem by sub-dividing rare words into relatively less rare sub-words. Such sub-division is also learned from large corpora of data. All the pre-trained models used in this paper used sub-word tokenization techniques. CodeBERT and GraphCodeBERT used RoBERTa tokenizer~\cite{liu2019roberta}, CodeGPT used GPT tokenizer~\cite{radford2019language}, and PLBART used sentence-piece tokenizer~\cite{kudo-richardson-2018-sentencepiece}. The use of such tokenizers strips away the burden of identifier abstraction. Our investigation shows that, in some cases, pre-trained models perform better with concrete tokens than abstract tokens (see \cref{tab:tokenization-discussion} for detailed result). Thus, we champion using input and outputs with concrete tokens when a pre-trained model is used. 
\section{Related Works}
\label{sec:related}

\subsection{Automatic Code Change}
\label{subsec:auto_code_change}
 
There are a lot of research efforts to capture repetitiveness of developers' way of editing source code. These researches show the potential of 
automatic refactoring~\citep{ge2012reconciling, raychev2013refactoring}, boilerplate code~\cite{meng2015does} etc. 
These research efforts include (semi-)automatic tools involving  traditional program analysis techniques (\eg clone detection, dependency analysis, graph matching)~\citep{ robbes2008example, meng2013lase}. Other research direction aims at learning source code edit from previous edits and applying those edit patterns in similar context~\citep{ray2014uniqueness,nguyen2013study}. 
Some of these efforts targets very specific code changes; 
{For example, Nguyen~\etal~\cite{nguyen2010graph} proposed a graph-matching-based approach for automatically updating API usage.
Tansey~\etal~\cite{tansey2008annotation} semantic preserving transformation of java classes for automated refactoring.
Other directions of works address more general-purpose code change learned from open source repositories~\citep{tufano2019learning, chakraborty2020codit}. Such approaches target solving automated code editing tasks in a data-driven approach, and the edit patterns are learned from example changes.} In this research, we also investigated general purpose source code changes in the wild. More closely to \tool, Rolim~\etal~\cite{rolim2017learning}'s proposed technique constraints source code generation with additional input/output specification or test cases. Nevertheless, we argue that textual guidance could be a very good surrogate specification.

\subsection{NMT for Code Change Modeling}
NMT has been studied for past couple of years to learn automatic source code change modeling. Tufano~\etal~\cite{tufano2018nmt_bug_fix, tufano2019empirical, tufano2019learning} presented initial investigation of using NMT in learning general purpose code changes. Chakraborty~\etal~\cite{chakraborty2020codit} proposed a tree based hierarchical NMT model for general purpose source code change. Instead of viewing code as sequence of tokens, they first generated syntax tree by sampling from Context Free Grammar, and then another model to fill up the gaps for identifier. To reduce the search space, they performed scope analysis to search for suitable identifier. Chen~\etal~\cite{chen2019sequencer} proposed a copy mechanism based NMT model for APR where the input is the code before change along with the context, and the output is the code after change. Their work treated the input as uni-modal way where the whole code is one singe modality. In this work, we consider multi-modal way of modeling, where we isolate the code fragment that needs to be changed from its context and present that code fragment concatenated with context to the model. Lutellier~\etal~\cite{lutellier2020coconut} treated code needs to be changed and the context as two difference modalities and use separate encoders. However, our empirical evidence showed that using one encoder to encode all the modalities result in the best performance.
{More recently, Ding~\etal~\cite{ding2020patching} presented empirical evidence that instead of generating a whole code element (\ie context+change) of the target version, only generating the sequence of changes might perform better for code change modeling. 
Recent works~\citep{tarlow2020learning, yao2021learning} proposed models for generating such edit sequence. Such models may augment or outperform NMT based code editing -- we leave such investigation as future work.}

\subsection{Machine Learning for Source Code Analysis}

In recent years, Machine Learning, especially Deep Learning has been widely adopted across different area of software engineering due to Availability of large collection of source code in open source platforms (\eg GitHub, Bitbucket, \etc) 
Application of ML based source code analysis include bug detection in code~\citep{ray2016naturalness}, clone detection~\citep{ wang2020detecting}, code completion~\citep{parvez2018building}, vulnerability detection~\citep{chakraborty2020deep}, code summarization~\citep{ahmad2020summarization}, code translation~\citep{  xu2020tree2tree}, etc. Recent works also approached to learn general purpose transferable representation learning for source code, which can later be used for various source code related tasks~\cite{alon2019code2vec, feng2020codebert,  jiang2021cure}. {The approaches for learning such transferable representations can be broadly categorized in two ways. The first category of approaches (\eg Code2Vec~\cite{alon2019code2vec}) aims at learning explicit representation for tokens in the code. Another category of approaches (\eg CodeBERT~\cite{feng2020codebert}) transfers syntactic and semantic interaction between code components in the form of pre-trained models. In this approach, a model for a specific task is initialized with a general-purpose pre-trained model, trained to understand and generate code.} In this paper, we empirically found that such pre-trained models (PLBART) increase accuracy upto 248\% in patch generation.

\section{Threats to Validity}
\label{sec:threats}

\subsection{External Validity}
\label{threats/external}

\noindent\textbf{Bias in the dataset.} Both \sdata, and \mdata are collection of bug-fix commits, and thus there is a threat that these dataset may exhibit specific bias towards bug-fix patches. While the commits in these datasets are filtered and classified as bug fix commits, these changes are made by real developers as part of development life cycle. Unlike other bugfix datasets~\cite{just2014defects4j}, \sdata and \mdata do not isolate the bug. Thus, we conjecture that possibility of existence of any such bias is minimal. 

\smallskip 

\noindent\textbf{Noise in commit message.} We used commit message as a guidance for code editing. While previous research efforts~\cite{zhou2019devign, zhou2017automated} showed that commit messages are very useful to summarize the changes in a commit, other research efforts~\cite{gallaba2018noise, kim2011dealing} also elucidated noises present in the commit message. To mitigate this threat, we carefully chose the dataset we tested \tool on. The original authors~\cite{tufano2019empirical} of the the dataset reported that they carefully investigated the dataset and after manual investigation, they reported that 97.6\% of the commits in their datasets are true positive. Despite this threat, \tool's performance seems to improve with commit message as additional input.

\subsection{Construct Validity}

{

In general, developers write commit message after they edited the code, in theory, summarizing the edits they made. In this paper, we assumed an experimental setup where developer would write the summary before editing the code. Such assumption may pose a threat to the applicability of \tool in real world, since in some cases, the developer may not know what edits they are going to make prior to the actual editing. 
Regardless, we consider \tool as a proof-of-concept, where empirically we show that, if a developer had the idea of change in mind, that could help an automated code editor.}

\subsection{Internal Threat}
All Deep Learning based techniques are sensitive to hyper-parameters. Thus using a sub-optimal hyper-parameter can pose a threat to the validity of \tool, especially while comparing with other baselines. As we compared with other pre-trained models, we cannot really modify the architecture and dimensions of other pre-trained models. As for other hyper-parameters (\ie learning rate, batch size, \etc), we use the exact same hyper-parameters described by respective paper. Nevertheless, we open source out code and data for broader dissemination.

\section{Conclusion}
\label{sec:conclusion}


In this paper, we highlight that an automatic code edit tool should possess knowledge about the underlying programming language, in general. Also, it can benefit from additional information such as edit context and developers' intention expressed in natural language. 
To that end, we design, present, and evaluate \tool -- a multi-modal NMT-based automated code editor. Our in-depth evaluation shows that \tool improves code-editing by leveraging knowledge about programming language through pre-training. In addition, we showed that leveraging additional modalities of information could benefit the source code editor. Our empirical evaluation reveals some critical lessons about the design choices of building an automated code editor that we believe will guide future research in automatic code editing. 
\section*{Acknowledgement}
This work is supported in part by NSF grants SHF-2107405, SHF-1845893, IIS-2040961, IBM, and VMWare. 
Any opinions, findings, conclusions, or recommendations expressed herein are those of the authors and do not necessarily reflect those of the US Government, NSF, IBM or VMWare.

\bibliographystyle{IEEEtran}
\bibliography{main}
\balance
\end{document}